\documentclass{PoS}
\bibliography{my-bib-database}
\usepackage{color}
\usepackage{wrapfig,lipsum,booktabs}
\def\1{\c{c}}
\def\2{\c{C}}
\def\3{\.{I}}
\def\4{\"{a}}
\def\5{{\i}}
\def\6{$\beta$}
\def\7{\"{o}}
\def\8{\"{O}}
\def\9{\c{s}}
\def\0{\c{S}}
\def\*{\"{u}}
\def\?{\"{U}}
\def\;{\u{g}}
\def\:{\u{G}}

\title{Detection of an Unidentified Extended Gamma-ray Source Close to the Galactic Supernova Remnant 3C 400.2}

\ShortTitle{Detection of an Unidentified Extended Gamma-ray Source Close to 3C 400.2}

\author{\speaker{T\*l\*n Ergin}
        TUBITAK Space Technologies Research Institute, Ankara, Turkey\\
        E-mail: \email{tulun.ergin@tubitak.gov.tr}}

\author{Aytap Sezer,
        Avrasya University, Trabzon, Turkey}

\author{Ryo Yamazaki,
        Aoyama Gakuin University, Fuchinobe, Japan}
          
\author{Hidetoshi Sano,
       Nagoya University, Nagoya, Japan}

\author{Yasuo Fukui,
       Nagoya University, Nagoya, Japan}

\author{Shuta Tanaka,
       Konan University, Kobe, Japan}
  
\abstract{A new extended gamma-ray source (PS J1934.5+1845) was detected with a significance of $\sim$13$\sigma$ at a location of 1$^{\circ}\!\!$.83 away from the radio location of the Galactic supernova remnant 3C 400.2 using about 9 years of Fermi-LAT data. The 68\% containment radius of PS J1934.5+1845's extension was found to be 0$^{\circ}\!\!$.61 and PS J1934.5+1845 is showing a power-law type spectrum with a spectral index of $\sim$2.38. In this presentation we will summarize the gamma-ray analysis methods and report on the analysis results related to the extension and spectrum of PS J1934.5+1845.}

\FullConference{7th Fermi Symposium 2017\\
		15-20 October 2017\\
		Garmisch-Partenkirchen, Germany}
		
\begin{document}

\vspace{-0.5cm}
\section{Introduction}
\vspace{-0.3cm}
While searching for gamma-ray emission from the mixed-morphology (MM; \cite{Rho98}) supernova remnant (SNR) 3C 400.2 (G53.6-2.2) \cite{Er17}, we detected a new extended unidentified gamma-ray source which we named as PS J1934.5+1845 after its best-fitted location. We start this paper by shortly explaining the analysis of {\it Fermi}-LAT data in Section 2. In Section 3, we summarize the analysis results for PS J1934.5+1845 and present our conclusions in Section 4. 

\vspace{-0.5cm}
\section{Data Reduction \& Analysis}
\vspace{-0.3cm}
Using the analysis toolkit \texttt{fermipy}\footnote{http://fermipy.readthedocs.io/en/latest/index.html} we analyzed {\it Fermi}-LAT gamma-ray data from the time period of 2008-08-04 to 2017-02-06 for events within the energy range of 200 MeV - 300 GeV. Using \texttt{gtselect} of {\it Fermi} Science Tools (FST)\footnote{https://fermi.gsfc.nasa.gov/ssc/data/analysis/software/}, we selected {\it Fermi}-LAT Pass 8 `Source' class and front+back type events, which come from zenith angles smaller than 90$^{\circ}$ and  from within a circular region of interest (ROI) with a radius of 30$^{\circ}$ centered at the radio position of 3C 400.2. The maximum likelihood fitting method \cite{Ma96} was employed on the spatially and spectrally binned data and used the instrument function P8R2$_{-}$SOURCE$_{-}\!\!$V6. The gamma-ray background model contains Galactic diffuse sources ({\it gll$_{-}$iem$_{-}\!$v6.fits}) and isotropic sources ({\it iso$_{-}$P8R2$_{-}$SOURCE$_{-}\!\!$V6$_{-}\!$v06.txt}). It also includes all point-like and extended sources from the 3rd {\it Fermi}-LAT source catalog (3FGL) \cite{Ac15} located within a 15$^{\circ}\times$15$^{\circ}$ region centered at the ROI center. Freed normalization parameters of sources that are within 3$^{\circ}$ of ROI center. Freed all parameters of the diffuse Galactic emission and the isotropic component. All sources with TS $>$ 10 are set free and all sources with TS $<$ 10 are fixed. 

The initial TS\footnote{Test statistics (TS) indicate that the null hypothesis (maximum likelihood value for a model without an additional source) is incorrect.} map was produced for a 10$^{\circ}\times$10$^{\circ}$ analysis region using this model showing gamma-ray excess close to the location of 3C 400.2 and at other locations within the analysis region. In order to obtain the exact TS value and location of the excess regions, we used an iterative source-finding algorithm in \texttt{fermipy}, called \texttt{find$_{-}$sources}:
\begin{itemize}
\vspace{-0.2cm}
\item Takes peak detection on a TS map to find new source candidates. The algorithm identified peaks with a significance threshold value higher than 3$\sigma$ and taking an angular distance of at least 1$^{\circ}\!\!$.5 from a higher amplitude peak in the map. 
\vspace{-0.2cm}
\item It orders the peaks by their TS values and adds a source at each peak starting from the highest TS peak. 
\vspace{-0.2cm}
\item Then it sets the source position by fitting a 2D parabola to the log-likelihood surface around the peak maximum.
\vspace{-0.2cm} 
\item After adding each source, having a significance value above 5$\sigma$, it re-fits the spectral parameters of that source. 
\vspace{-0.2cm}
\item With this algorithm, we identified nine new sources within the analysis region. 
\vspace{-0.2cm}
\item The algorithm also listed the sources with a significance between 3$\sigma$ and 5$\sigma$. One of these sources, PS J1938.6+1722, was within the 95\% confidence radius of the 3C 400.2 position. So, the name of PS J1938.6+1722 is used for 3C 400.2. 
\vspace{-0.2cm}
\end{itemize}

As a next step we analyzed the closer vicinity of 3C 400.2 by taking an analysis region of 5$^{\circ}\times$5$^{\circ}$. We followed the same analysis procedure as described for the 10$^{\circ}\times$10$^{\circ}$ analysis region, except that we added two sources into the background model: PS J1934.5+1845, being the only bright gamma-ray source having a significance above 5$\sigma$ within the analysis region, and PS J1938.6+1722, representing 3C 400.2. 

\begin{figure}
\centering
\includegraphics[width=0.9\textwidth]{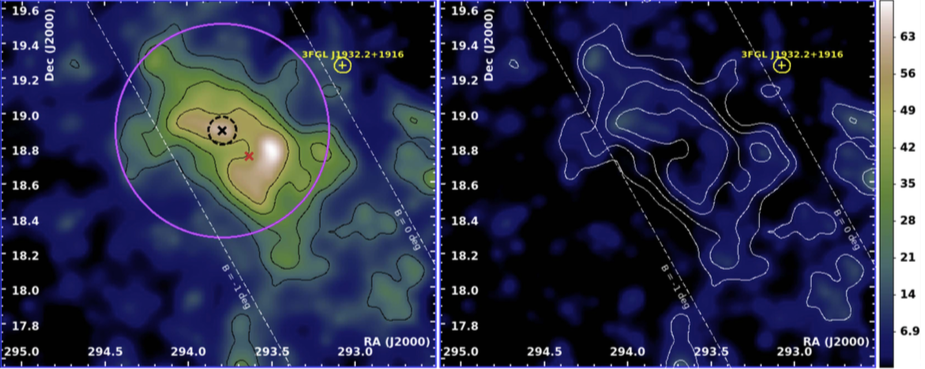}
\vspace{-0.2cm}
\caption{ \footnotesize{TS map of the gamma-ray emission, where PS J1934.5+1845 was not included in the background model (left panel). The initial spatial model of the source was taken as point-like, where its location was chosen as the red cross shown in the left panel. The best-fit position of the point-like source is shown with a black cross, and its positional error at a 95\% confidence level is shown as a black dashed circle. The magenta solid circle represents the best-fit extension found using a radial Gaussian model centered around the best-fit position of PS J1934.5+1845. The right panel shows the gamma-ray TS map, where PS J1934.5+1845 was included as a point-like gamma-ray source in the background model. The black and white contours on both panels are the levels of the TS values, which are 16, 25, 36, 49, 69, and a gamma-ray source from 3FGL \cite{Ac15} is shown in yellow, together with its positional error circle. The white dashed lines correspond to the Galactic latitudes of -1$^{\circ}$ and 0$^{\circ}$.}}
\label{figure_1}
\vspace{-0.3cm}
\end{figure}

\vspace{-0.5cm}
\section{Analysis Results}

\begin{wrapfigure}{r}{0.5\textwidth} 
\vspace{-20pt}
\begin{center}
\includegraphics[width=0.5\textwidth]{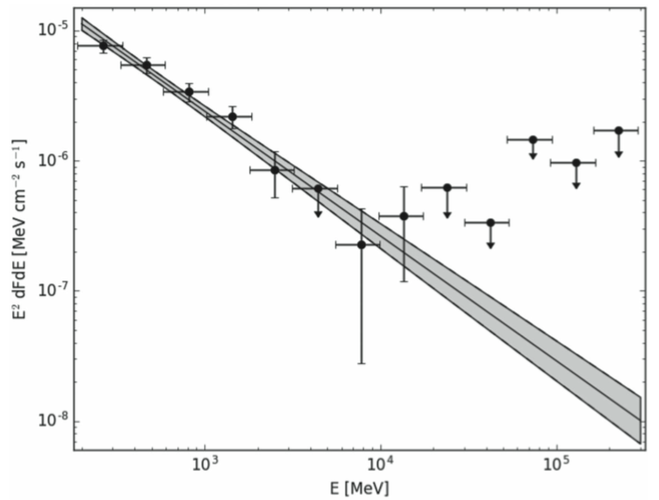}
\vspace{-23pt}
\caption{\footnotesize{Gamma-ray spectral energy distribution of PS J1934.5+1845 assuming it to be an extended source. The shaded region represents the model flux and its statistical errors obtained from fitting a PL-type spectrum to the given spectral data.}}
\label{figure_2}
\end{center}
\vspace{-25pt}
\end{wrapfigure}

\vspace{-0.3cm}
\subsection{Detection \& Localization} 
\vspace{-0.3cm}
PS J1934.5+1845, was detected at a location of 1$^{\circ}\!\!$.83 away from the radio location of 3C 400.2. The detection significance was found to be $\sim$13$\sigma$ (i.e., TS = 161), assuming PS J1934.5+1845 as a point-like source during the source search procedure. Using the \texttt{localize} method of \texttt{fermipy}, the best-fitted position for PS J1934.5+1845 was found to be R.A. (J2000) = 293$^{\circ}\!\!$.79 $\pm$ 0$^{\circ}\!\!$.08 and decl. (J2000) =18$^{\circ}\!\!$.90 $\pm$ 0$^{\circ}\!\!$.08 (R.A. (J2000) = 19h 35m 10s.32, decl. (J2000) = 18$^{\circ}$ 54$'$ 07$''\!\!$.20).

We checked 3FGL \cite{Ac15} and the 3rd catalog of hard Fermi-LAT sources (3FHL) \cite{Ac17} to find possible counterparts for PS J1934.5+1845. Figure \ref{figure_1} shows the TS map of PS J1934.5+1845. {\it Fermi}-LAT sources from 3FGL \cite{Ac15} as yellow markers. We could not find any counterparts for SourceA in 3FGL catalog.   

\vspace{-0.3cm}
\subsection{Extension Measurements}
\vspace{-0.3cm}
We used two models to parameterize the extended gamma- ray emission morphology of PS J1934.5+1845, disk and radial Gaussian models, where the width and location of the centers of the models were calculated by the \texttt{extension} method of \texttt{fermipy}. To detect the extension of a source, we used the TS of the extension (TS$_{ext}$) parameter, which is the likelihood ratio comparing the likelihood for being a point-like source (L$_{pt}$) to a likelihood for an existing extension (L$_{ext}$), TS$_{ext}$ = 2log (L$_{ext}$/L$_{pt}$). The 'Extension Width' which is the 68\% containment radius of the extension model (R$_{68}$), was found to be 0$^{\circ}\!\!$.6107 + 0$^{\circ}\!\!$.0866 - 0$^{\circ}\!\!$.1141, with a TS$_{ext}$ value of $\sim$40 for the radial Gaussian model. As an extended source, PS J1934.5+1845 was detected with a significance of $\sim$13$\sigma$ (i.e. TS = 168). 

\vspace{-0.3cm}
\subsection{Spectral Measurements}
\vspace{-0.3cm}
The spectrum was fit to PL model, where the spectral index is found to be $\Gamma$ = 2.98 $\pm$ 0.09. The total photon flux and energy flux was found to be (2.99 $\pm$ 0.31) $\times$ 10$^{-9}$ photons cm$^{-2}$ s$^{-1}$ and (1.20 $\pm$ 0.11) $\times$ 10$^{-6}$ MeV cm$^{-2}$ s$^{-1}$, respectively, for the point-like source model having a PL-type spectrum. Assuming a PL-type spectrum for this extended source, we obtained $\Gamma$ = 2.38 $\pm$ 0.07, and the total photon flux and energy flux of PS J1934.5+1845 was found to be (2.54 $\pm$ 0.23) $\times$ 10$^{-8}$ photons cm$^{-2}$ s$^{-1}$ and (1.74 $\pm$ 0.16) $\times$ 10$^{-5}$ MeV cm$^{-2}$ s$^{-1}$, respectively. The spectrum for the extended emission of PS J1934.5+1845 is shown in Figure \ref{figure_2}.

\vspace{-0.3cm}
\section{Conclusion \& Outlook}
\vspace{-0.3cm}
We detected a new source, PS J1934.5+1845, at about 1$^{\circ}\!\!$.8 away from 3C 400.2, having a significance of $\sim$13$\sigma$. PS J1934.5+1845 was found to have a radial Gaussian type extension with a radius of $\sim$0$^{\circ}\!\!$.61. We investigated this source as a part of our analysis, due to the possibility that it might have been contributing to the gamma-ray emission of 3C 400.2. After fitting the extension and checking the TS value of 3C 400.2, we found out that the emission of PS J1934.5+1845 is not directly affecting the gamma-ray emission of 3C 400.2. The gamma-ray emission of PS J1934.5+1845 needs to be further investigated by modeling the SED in order to understand whether the dominating gamma-ray emission mechanism is leptonic or hadronic. 

As a next step, we plan for multi-waveband observations (radio, optical, and X-rays). In addition, we will re-analyze the gamma-ray data  to do more variability checks and investigate the energy-dependent source morphology.


\end{document}